# A candidate super-Earth planet orbiting near the snow line of Barnard's star

I. Ribas[1,2], M. Tuomi[3], A. Reiners[4], R. P. Butler[5], J. C. Morales[1,2], M. Perger[1,2], S. Dreizler[4], C. Rodríguez-López[6], J. I. González Hernández[7,8], A. Rosich[1,2], F. Feng[3], T. Trifonov[9], S. S. Vogt[10], J. A. Caballero[11], A. Hatzes[12], E. Herrero[1,2], S. V. Jeffers[4], M. Lafarga[1,2], F. Murgas[7,8], R. P. Nelson[13], E. Rodríguez[6], J. B. P. Strachan[13], L. Tal-Or[4,14], J. Teske[5,15], B. Toledo-Padrón[7,8], M. Zechmeister[4], A. Quirrenbach[16], P. J. Amado[6], M. Azzaro[17], V. J. S. Béjar[7,8], J. R. Barnes[18], Z. M. Berdiñas[19], J. Burt[20], G. Coleman[21], M. Cortés-Contreras[11], J. Crane[22], S. G. Engle[23], E. F. Guinan[23], C. A. Haswell[18], Th. Henning[9], B. Holden[10], J. Jenkins[19], H. R. A. Jones[3], A. Kaminski[16], M. Kiraga[24], M. Kürster[9], M. H. Lee[25], M. J. López-González[6], D. Montes[26], J. Morin[27], A. Ofir[28], E. Pallé[7,8], R. Rebolo[7,8,29], S. Reffert[16], A. Schweitzer[30], W. Seifert[16], S. A. Shectman[22], D. Staab[18], R. A. Street[31], A. Suárez Mascareño[32,7], Y. Tsapras[33], S. X. Wang[5], G. Anglada-Escudé[13,6]

[1] Institut de Ciències de l'Espai (ICE, CSIC), Campus UAB, C/Can Magrans s/n, 08193 Bellaterra, Spain
[2] Institut d'Estudis Espacials de Catalunya (IEEC), 08034 Barcelona, Spain
[3] Centre for Astrophysics Research, University of Hertfordshire, College Lane, AL10 9AB, Hatfield, United Kingdom
[4] Institut für Astrophysik Göttingen, Georg-August-Universität Göttingen, Friedrich-Hund-Platz 1, 37077 Göttingen, Germany
[5] Department of Terrestrial Magnetism, Carnegie Institution for Science, 5241 Broad Branch Road NW, Washington DC 20015, USA
[6] Instituto de Astrofísica de Andalucía (IAA, CSIC), Glorieta de la Astronomía 1, 18008 Granada, Spain
[7] Instituto de Astrofísica de Canarias (IAC), E-38205 La Laguna, Tenerife, Spain

[8] Universidad de La Laguna (ULL), Departamento de Astrofísica, E-38206 La Laguna, Tenerife, Spain
[9] Max-Planck-Institut für Astronomie, Königstuhl 17, D-69117 Heidelberg, Germany
[10] UCO/Lick Observatory, University of California at Santa Cruz, 1156 High Street, Santa Cruz, CA 95064, USA
[11] Centro de Astrobiología, CSIC-INTA, ESAC campus, Camino bajo del castillo s/n, 28692, Villanueva de la Cañada, Madrid, Spain
[12] Thüringer Landessternwarte, Sternwarte 5, D-07778 Tautenburg, Germany
[13] School of Physics and Astronomy, Queen Mary University of London, 327 Mile End Rd, E1 4NS London, United Kingdom
[14] School of Geosciences, Raymond and Beverly Sackler Faculty of Exact Sciences, Tel-Aviv University, Tel Aviv, 6997801, Israel
[15] Hubble Fellow
[16] Landessternwarte, Zentrum für Astronomie der Universität Heidelberg, Königstuhl 12, 69117 Heidelberg, Germany
[17] Centro Astronómico Hispano-Alemán (CSIC-MPG), Observatorio Astronómico de Calar Alto, Sierra de los Filabres, E-04550 Gérgal, Almería, Spain
[18] School of Physical Sciences, The Open University, Robert Hooke Building, Walton Hall, MK7 6AA Milton Keynes, United Kingdom
[19] Departamento de Astronomía, Universidad de Chile, Camino El Observatorio, 1515 Las Condes, Santiago, Chile
[20] Kavli Institute, Massachusetts Institute of Technology, 77 Massachusetts Avenue, Cambridge, MA 02139, USA
[21] Physikalisches Institut, Universität Bern, Silderstrasse 5, 3012 Bern, Switzerland
[22] The Observatories, Carnegie Institution for Science, 813 Santa Barbara Street, Pasadena, CA, 91101, USA
[23] Department of Astrophysics & Planetary Science, Villanova University, 800 Lancaster Ave., Villanova, PA 19085, USA
[24] Warsaw University Observatory, Aleje Ujazdowskie 4, 00-478 Warszawa, Poland
[25] Department of Earth Sciences and Department of Physics, The University of Hong Kong, Pokfulam Road, Hong Kong
[26] Dep. de Física de la Tierra Astronomía y Astrofísica & UPARCOS-UCM (Unidad de Física de Partículas y del Cosmos de la UCM), Facultad de Ciencias Físicas, Universidad Complutense de Madrid, Av. Complutense s/n, 28040 Madrid, Spain
[27] Laboratoire Univers et Particules de Montpellier, Université de Montpellier, CNRS Place Eugène Bataillon, F-34095 Montpellier, France
[28] Department of Earth and Planetary Sciences, Weizmann Institute of Science, 234 Herzl St., PO Box 26, Rehovot 7610001, Israel
[29] Consejo Superior de Investigaciones Científicas (CSIC)
[30] Hamburger Sternwarte, Universität Hamburg, Gojenbergsweg 112, 21029 Hamburg, Germany
[31] Las Cumbres Observatory Global Telescope Network, 6740 Cortona Drive, suite 102, Goleta, CA 93117, USA
[32] Observatoire Astronomique de l'Université de Genève, 1290 Versoix, Switzerland
[33] Zentrum für Astronomie der Universität Heidelberg, Astronomisches Rechen-Institut, Mönchhofstr. 12-14, 69120 Heidelberg, Germany

**At a distance of 1.8 parsecs[1], Barnard's star (Gl 699) is a red dwarf with the largest apparent motion of any known stellar object. It is the closest single star to the Sun, second only to the $\alpha$ Centauri triple stellar system. Barnard's star is also among the least magnetically active red dwarfs known[2,3] and has an estimated age older than our Solar System. Its properties have made it a prime target for planet searches employing techniques such as radial velocity[4,5,6], astrometry[7,8], and direct imaging[9], all with different sensitivity limits but ultimately leading to disproved or null results. Here we report that the combination of numerous measurements from high-precision radial velocity instruments reveals the presence of a low-amplitude but significant periodic signal at 233 days. Independent photometric and spectroscopic monitoring, as well as the analysis of instrumental systematic effects, show that this signal is best explained as arising from a planetary companion. The candidate planet around Barnard's star is a cold super-Earth with a minimum mass of 3.2 Earth masses orbiting near its snow-line. The combination of all radial velocity datasets spanning 20 years additionally reveals a long-term modulation that could arise from a magnetic activity cycle or from a more distant planetary object. Because of its proximity to the Sun, the proposed planet has a maximum angular separation of 220 milli-arcseconds from Barnard's star, making it an excellent target for complementary direct imaging and astrometric observations.**

Barnard's star is the second closest red dwarf to the Solar System, after Proxima Centauri, and thus an ideal target to search for exoplanets with potential for further characterisation[10]. Its very low X-ray flux, lack of H$\alpha$ emission, low chromospheric emission indices, slow rotation rate, slightly sub-solar metallicity, and membership of the thick disc kinematic population indicate an extremely low magnetic activity level and suggest an age older than the Sun. Because of its apparent brightness and very low variability, Barnard's star is often regarded as a benchmark for intermediate M-type dwarfs. Its basic properties are summarized in Table 1.

An early analysis of archival radial velocity datasets of Barnard's star up to 2015 indicated the presence of at least one significant signal with a period of ~230 days but with rather poor sampling. To elucidate its presence and nature we undertook an intensive monitoring campaign with the CARMENES spectrometer[11], collecting precise radial velocity measurements on every possible night during 2016-2017, and we obtained overlapping observations with the ESO/HARPS and HARPS-N instruments. The combined Doppler monitoring effort of Barnard's star, including archival and newly acquired observations, resulted in 771 radial velocity epochs (nightly averages) with typical individual precisions of 0.9 to 1.8 m $s^{-1}$, obtained over a timespan exceeding 20 years from seven different facilities and yielding eight independent datasets (ED Table 1).

While each dataset is internally consistent, relative offsets may be present because of uncertainties in the absolute radial velocity scale. The analysis considered a zero-point value and a noise term (jitter) for each dataset as free parameters to be optimized simultaneously with the planetary models, and a global linear trend. We used several independent fitting methods to ensure the reliability of the results. The parameter space was scanned with hierarchical procedures (signals are identified individually and added recursively to the model) and multi-signal search approaches (fitting two or more signals at a time). Furthermore, we used the Systemic Console[12] to assess the sensitivity of the solutions to the datasets used, error estimates and eccentricities. Figure 1 and ED Figure 1 illustrate the detection of a signal at a period of 233 days with high statistical significance assuming white noise (p-value or false-alarm probability, FAP ~ $10^{-15}$) and also show evidence for a second, longer-period signal.

To assess the presence of the long-term modulation we considered an alternative method of determining the relative offsets by directly averaging radial velocity differences within defined time intervals for overlapping observations. All datasets were subsequently "stitched" together into a single radial velocity time-series. These combined measurements indicate long-term variability consistent with a signal at a period greater than 6000 days. We thus performed additional fits leaving the relative offsets as free parameters and assuming two signals, one with a prior allowing only periods > 4000 days. The model fit converges to two periodic signals at 233 days and ~6600 days, and has comparable likelihood ($\Delta \ln L < 5$) to the one obtained by manually "stitching" the datasets. We conclude that the significance of the 233-day signal remains unaltered irrespective of the model used for the long-term variability, and also that the long-term variability is significant.

Stellar activity is known to produce periodic radial velocity modulations that could be misinterpreted as arising from planetary companions. Rotation period values of 130 days and 148.6 days have been reported for Barnard's star respectively from photometry[13] and from spectroscopic indices[3]. We analysed data from long-term monitoring in photometry and spectroscopy, the latter being Hα and Ca II H&K chromospheric fluxes measured from the spectra used for radial velocity determination. Periodograms are shown in Figure 2. The photometric time-series yields a statistically significant signal with a period of 144 days, the Hα measurements present a complex periodogram with a highly significant main peak at 133 days, and the Ca II H&K chromospheric index shows significant periodicity at 143 days. All these values can be tentatively associated to the stellar rotation period, which we hereby estimate to be 140±10 days. Furthermore, two of the activity tracers suggest the existence of long-term variability. The analysis rules out stellar activity periodicities in the neighbourhood of 230 days. Also, the 233-day signal in radial velocity increases significance mostly monotonically with time as additional observations are accumulated (ED Figure 2), which is suggestive of a deterministic Keplerian motion rather than the more stochastic nature of stellar activity variations.

Although stellar activity does not appear to be responsible for the periodic 233-day signal in radial velocity, it could affect the significance and determination of the model parameters. We therefore carried out a detailed study considering different models for correlated noise, based on Moving Averages (MA) and Gaussian Processes (GP). The MA models yield results comparable with the analysis assuming white noise and confirm the high statistical significance of the 233-day periodicity, with a FAP of $5\cdot10^{-10}$. The GP framework strongly reduces the signal significance, with a FAP no more significant than ~10%. However, GP models have been shown[14] to underestimate the significance of the signals, even in the absence of correlated noise.

Despite the degeneracies encountered with certain models, and after extensive testing (see Methods for further details), we conclude that the 233-day period signal in the radial velocities is best explained as arising from a planet with minimum mass of 3.2 Earth masses in a low-eccentricity orbit of 0.40 au semi-major axis. The median parameter values from our analysis are provided in Table 1 and ED Table 2, while Figure 3 shows the models of the radial velocities. Standard Markov chain Monte Carlo (MCMC) procedures were used to sample the posterior distribution. The MCMC analysis yields a secular trend significantly different from zero. Both the trend and the long-term modulation could be related to a stellar activity cycle (as photometric and spectroscopic indicators may suggest) but the presence of an outer planet cannot be ruled out. In the latter case, the fit would suggest an object of $\gtrsim 15$ Earth masses, in an orbit with ~4 au semi-major axis. The orbital period is compatible with that claimed by ref. 6 from an astrometric long-term study, but the Doppler amplitude is inconsistent, unless the orbit is nearly face-on. On the other hand, the induced nonlinear astrometric signature over ~5 yr would be up to 3 milliarcseconds, making it potentially detectable with the Gaia mission.

ED Figure 1 shows that some marginally significant signals may be present in the residuals of the two-signal model (e.g., at 81 d), but current evidence is inconclusive. We can, however, set stringent limits on the exoplanet detectability in close-in orbits around Barnard's star. Our analysis is sensitive to planets with minimum masses 0.7 and 1.2 Earth masses at respective orbital periods of 10 and 40 days, which correspond to the inner and outer optimistic habitable zone limits[15]. Barnard's star seems to be devoid of Earth-mass planets and larger in hot and temperate orbits, which stands in contrast with the seemingly high occurrence of planets in close-in orbits around M-type stars found by the Kepler mission[16,17].

The proximity of Barnard's star and the relatively large orbital separation makes the system ideal for astrometric detection. The *Gaia* and *HST* missions can reach an astrometric accuracy of 0.03 mas[18,19]. Depending on the orbital inclination they could detect the planet signal or set a constraining mass upper limit[20]. Also, for the calculated orbital separation the contrast ratio between the planet and the star in reflected light is of the order of a few times $10^{-9}$ depending on the adopted values of the geometric albedo and orbital inclination. This is beyond the capabilities of current imaging instrumentation by three orders of magnitude. However, the maximum apparent separation is 220 mas, which should be within reach of planned direct imaging instruments for the next decade[21], potentially revealing a wealth of information.

The candidate planet Barnard's star b lies almost exactly at the expected position of the snow-line of the system, located at about 0.4 au (ref. 22). It has long been suggested that this region might provide a favourable location for forming planets[23,24], with super-Earths being the most common planets formed around low-mass stars[25]. Recent models incorporating dust coagulation, radial drift, and planetesimal formation via the streaming instability support this idea[26]. Although this has yet to be shown to be part of a general trend, observational evidence would significantly constrain theories of planetary migration[27].

The long-term intensive monitoring of Barnard's star and the precision of the measurements, gathering data from all precise high-resolution spectrometers in operation, pushes the limits of the radial velocity technique into a new regime of parameter space, namely super-Earth type planets in cool orbits. This provides a bridge with the microlensing technique, which has traditionally been the only probe to explore the occurrence of small planets in orbits around the snow-line[28,29].

**Acknowledgments.** The results are based on observations made with the CARMENES instrument at the 3.5-m telescope of the Centro Astronómico Hispano-Alemán de Calar Alto (CAHA, Almería, Spain), funded by the German Max-Planck-Gesellschaft (MPG), the Spanish Consejo Superior de Investigaciones Científicas (CSIC), the European Union, and the CARMENES Consortium members, the 90-cm telescope at the Sierra Nevada Observatory (Granada, Spain) and the 40-cm robotic telescope at the SPACEOBS observatory (San Pedro de Atacama, Chile), both operated by the Instituto de Astrofísica de Andalucía (IAA), and the 80-cm Joan Oró Telescope (TJO) of the Montsec Astronomical Observatory (OAdM), owned by the Generalitat de Catalunya and operated by the Institute of Space Studies of Catalonia (IEEC). This research was supported by the following institutions, grants and fellowships: Spanish MINECO ESP2016-80435-C2-1-R, ESP2016-80435-C2-2-R, AYA2016-79425-C3-1-P, AYA2016-79245-C3-2-P, AYA2016-79425-C3-3-P, AYA2015-69350-C3-2-P, ESP2014-54362-P, AYA2014-56359-P, RYC-2013-14875; Generalitat de Catalunya/CERCA programme; Fondo Europeo de Desarrollo Regional (FEDER); German Science Foundation (DFG) Research Unit FOR2544, project JE 701/3-1; STFC Consolidated Grants ST/P000584/1, ST/P000592/1, ST/M001008/1; NSF AST-0307493; Queen Mary University of London Scholarship; Perren foundation grant; CONICYT-FONDECYT 1161218, 3180405; Swiss National Science Foundation (SNSF); Koshland Foundation and McDonald-Leapman grant; and NASA Hubble Fellowship grant HST-HF2-51399.001.

**Author Contribution** I.R. led the CARMENES team and the TJO photometry, organized the analysis of the data, and wrote most of the manuscript. M.T. performed the initial radial velocity analysis and, with, J.C.M., M.P., S.D., A.Ro., F.F., T.T., S.S.V., A.H., A.K., S.S.V., J.J., and A.S.M. participated in the analysis of radial velocity data using various methods. A.Re. co-led the CARMENES team. R.P.B. led the HIRES/PFS/APF team and reprocessed the UVES data. C.R.-L. coordinated the acquisition and analysis of photometry. J.I.G.H., R.R., A.S.M, and B.T.-P. acquired HARPS-N data and measured chromospheric indices from all spectroscopic datasets. T.T. and M.H.L., studied the dynamics. S.S.V. co-led the HIRES/APF teams. J.A.C. is responsible for the CARMENES instrument and, with A.S. and M.C.-C., determined the stellar properties. E.H., F.M., E.R., J.B.P.S., S.G.E., E.F.G., M.Ki., and M.J.L.-G. participated in the photometric monitoring. S.V.J. contributed to the analysis of activity and to the preparation of the manuscript. M.L. calculated the cross-correlation function parameters of CARMENES spectra. R.N. participated in the discussion of planet formation implications. A.Q. and P.J.A. are principal investigators of CARMENES. M.A., V.J.S.B., T.H., M.Ku., D.M., E.P., S.R., and W.S. are members of the CARMENES consortium. L.T.-O. calibrated the CARMENES data and carried out calculations of astrometric detection. M.Z. reduced the CARMENES data. J.T., J.B., J.D.C., B.H., S.A.S., and S.X.W. participated in the acquisition and discussion of HIRES, PFS and APF data. J.R.B., G.C., C.A.H, J.J., H.R.A.J, J.M., A.O., D.S., R.A.S., and Y.T. participated in the RedDots2017 collaboration. Z.M.B. participated in the discussion of instrument systematics. G.A.-E. organized the collaboration, coordinated the RedDots2017 campaign, organized and performed analyses and participated in the manuscript writing. All authors were given the opportunity to review the results and comment on the manuscript.

**Author Information** Reprints and permissions information is available at www.nature.com/reprints. The authors declare no competing interests. Correspondence and requests for materials should be addressed to iribas@ice.cat.

## Tables

| Star parameter | Value |
|---|---|
| Spectral type | M3.5 V |
| Mass ($M_{\odot}$) | 0.163±0.022 |
| Radius ($R_{\odot}$) | 0.178±0.011 |
| Luminosity ($L_{\odot}$) | 0.00329±0.00019 |
| Effective temperature (K) | 3278±51 |
| Rotation period (d) | 140±10 |
| Age (Ga) | 7–10 |
| **Planet parameter** | **Value** |
| | Barnard's star b |
| Orbital period (d) | $232.80^{+0.38}_{-0.41}$ |
| Radial velocity semi-amplitude (m $s^{-1}$) | 1.20±0.12 |
| Eccentricity | $0.32^{+0.10}_{-0.15}$ |
| Argument of periastron (deg) | $107^{+19}_{-22}$ |
| Mean longitude at BJD2455000.0 (deg) | 203±7 |
| Minimum mass ($M \sin i$; $M_{\oplus}$) | 3.23±0.44 |
| Orbital semi-major axis (au) | 0.404±0.018 |
| Irradiance (Earth units) | 0.0203±0.0023 |
| Equilibrium temperature (K) | ≲105±3 |
| Minimum astrometric semi-amplitude ($\alpha \sin i$; mas) | 0.0133±0.0013 |
| Angular separation (mas) | 221±10 |

**Table 1: Information on Barnard's star and its planet.**

We derive fundamental parameters of Barnard's star as in ref. 30. The luminosity is calculated from a well-sampled spectral energy distribution and the effective temperature is used to derive the stellar radius. The age interval is estimated by considering kinematic parameters, stellar rotation, and magnetic activity indicators. The parameters of the planet and their uncertainties are determined by calculating the median values and 68% credibility intervals of the distribution resulting from the MCMC run. The equilibrium temperature value is calculated assuming only external energy sources, with the upper limit corresponding to a null Bond albedo.

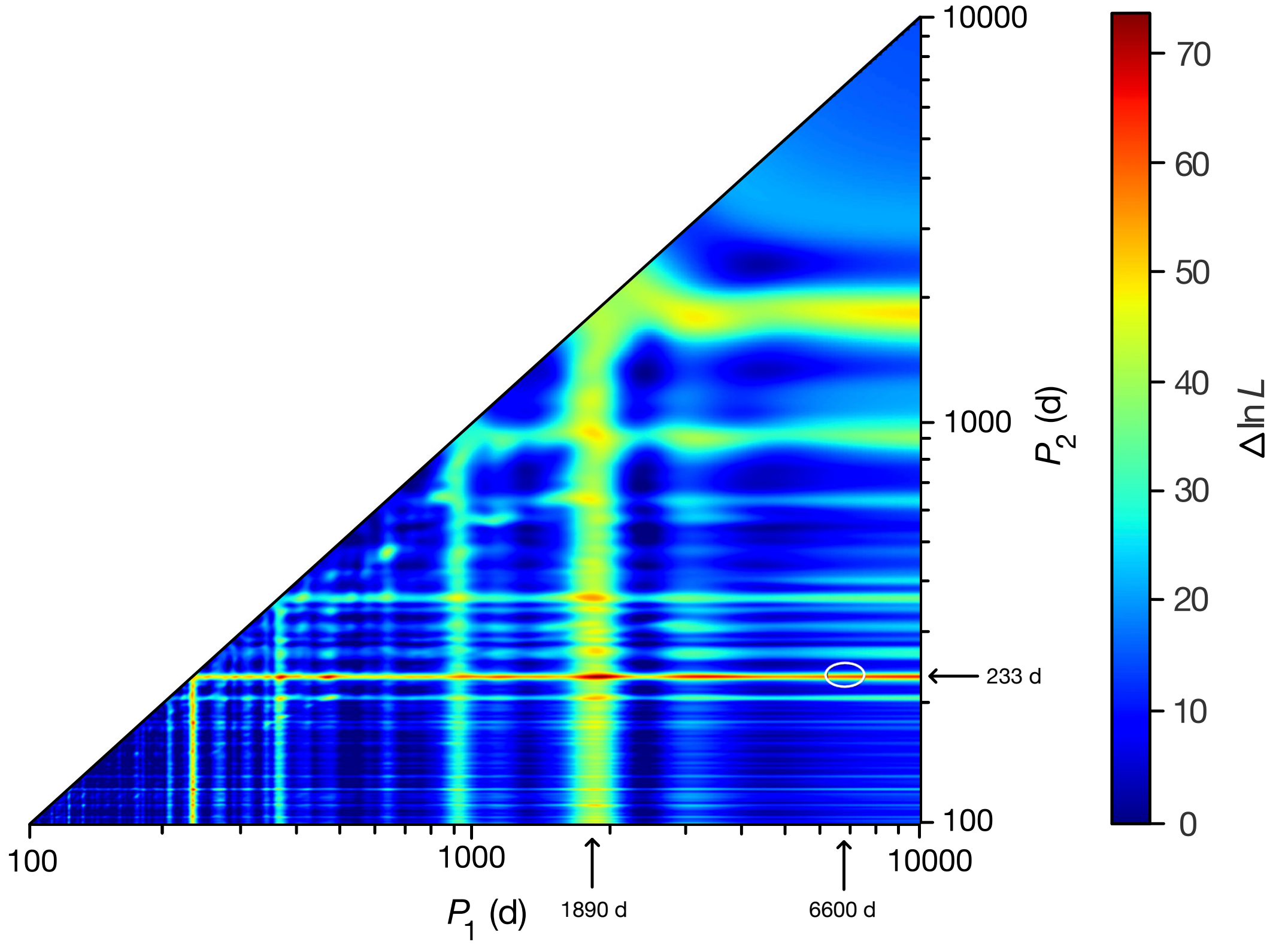

**Figure 1: Two-dimensional likelihood periodogram.** A multi-dimensional generalised Lomb-Scargle scheme assuming a white noise model was used to explore combinations of periods to fit the data. The colour scale shows the improvement of the logarithm of the likelihood function $\Delta \ln L$ as a function of trial periods. $\Delta \ln L > 18.1$ corresponds to a significant detection (FAP < 0.1%) for one signal, while two signals require $\Delta \ln L > 36.2$. The highest likelihood value corresponds to periods of 233 days and 1890 days ($\Delta \ln L = 71$), but any combination of 233 days with periods longer than 2500 days yields $\Delta \ln L > 65$ and thus are statistically equivalent. The proposed solution discussed in the text ($P_1$ = 233 d, $P_2$ = 6600 d) is indicated with a dashed ellipse.

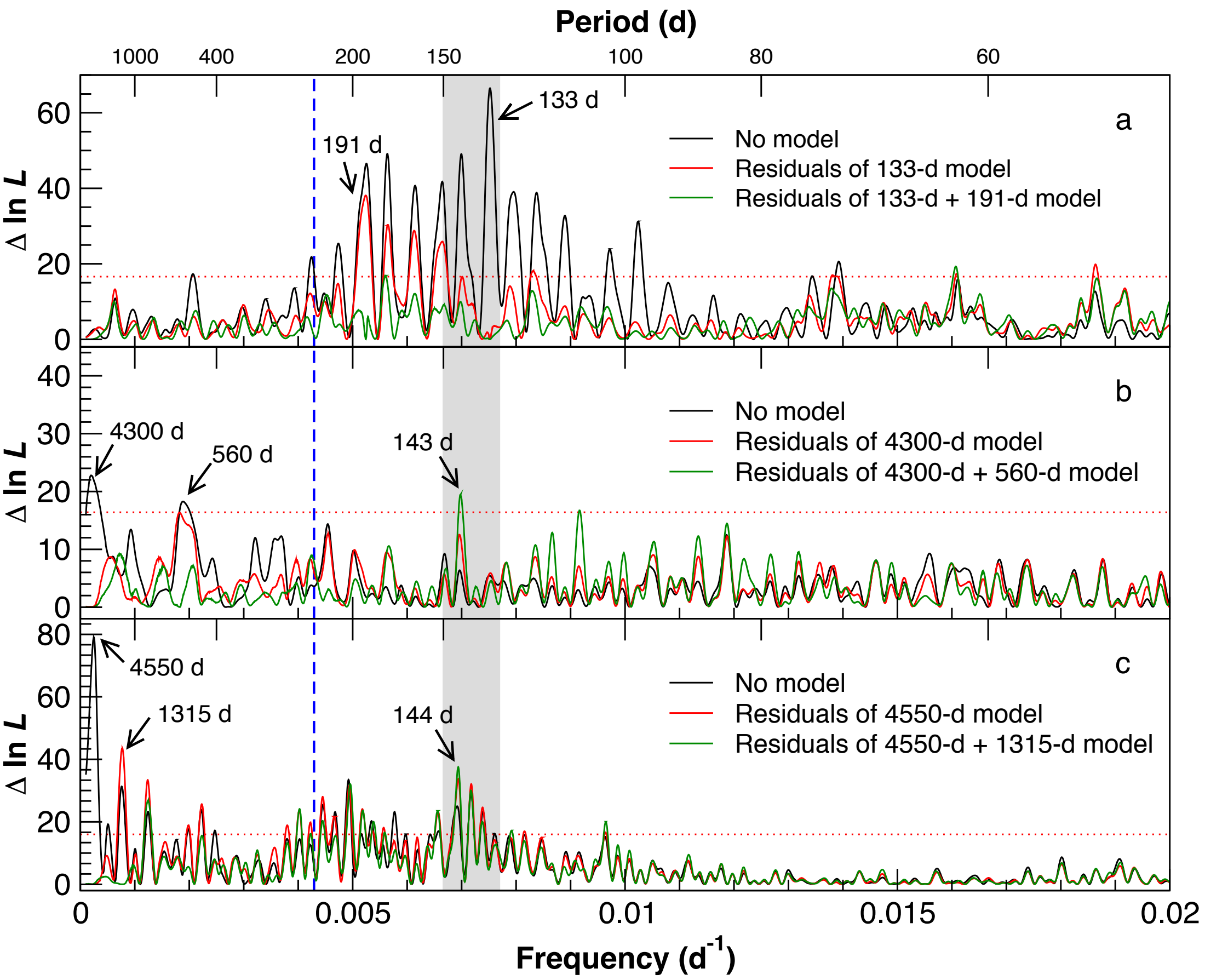

**Figure 2: Periodicities in stellar activity indicators.** The panels show periodograms of time series in the central flux of the Hα line (a), the emission in the Ca II H&K lines (b) and photometry (c). These indicators are associated to the presence of active regions on the stellar surface. Likelihood periodograms were obtained by including one signal at a time (sinusoids) as in the analysis of the radial velocities. The vertical dashed blue line indicates the location of the planetary signal from the radial velocity analysis, at a period of 233 days, while the dotted red line shows the FAP=0.1% detection threshold. The shaded region marks the most likely stellar rotation interval.

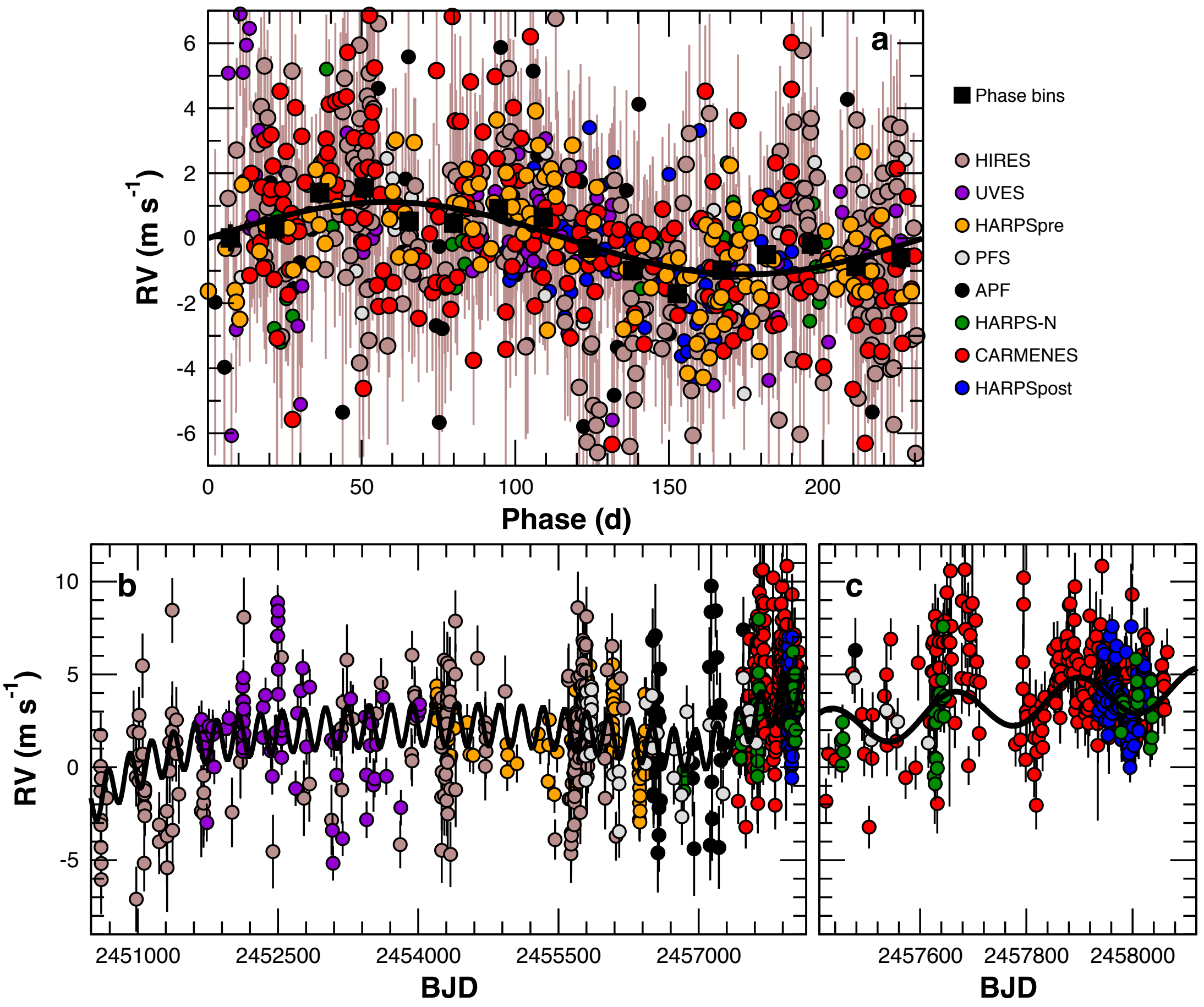

**Figure 3: Radial velocity time-series fits.** Panel a shows the phase-folded representation of the best-fitting 233-day circular orbit (black line) to the different sets (circles). The black squares represent the average velocity in 16 bins along the orbital phase. The lower panels show the time series of the radial velocity observations with the fitted model superimposed (b) and a blow-up of the time region around CARMENES observations (c). The model fit corresponds to a solution assuming two signals (one of them forced to P > 4000 days, for reasons discussed in the text). In all cases, 1σ error bars on the measurements are shown.

# METHODS

## Description of the individual radial velocity datasets

As in other recent low-amplitude exoplanet discoveries, combining information from several instruments (historical data and quasi-simultaneous monitoring) is central to unambiguously identifying significant periodicities in the data. The suite of instruments used for this study and relevant information on the observation time intervals, the number of epochs, and the references of the observational programs involved are provided in ED Table 1.

The HIRES, PFS, and APF datasets were obtained respectively with the HIRES spectrometer[31] on the Keck I 10-m telescope atop Mauna Kea in Hawaii, the Planet Finding Spectrometer (PFS) (ref. 32) on Carnegie's Magellan II 6.5-m telescope, and the Automated Planet Finder (APF) (ref. 33) on the 2.4-m telescope atop Mt. Hamilton at Lick Observatory. In all cases, radial velocities were calibrated by placing a cell of gaseous iodine in the converging beam of the telescope, just ahead of the spectrometer slit. The iodine superimposes a rich forest of absorption lines on the stellar spectrum over the 5000–6200 Å region, thereby providing a wavelength calibration and proxy for the point spread function (PSF) of the spectrometer. Once extracted, the iodine region of each spectrum is divided into 2 Å wide chunks, resulting in ~700 chunks for both the APF and HIRES, and ~800 for PFS. Each chunk produces an independent measure of the absolute wavelength, PSF, and Doppler shift, determined using the spectral synthesis technique described in ref. 34. The final reported Doppler velocity of each stellar spectrum is the weighted mean of the velocities of all the individual chunks. The final uncertainty of each velocity is the standard deviation of all chunk velocities about the weighted mean.

Further radial velocity measurements of Barnard's star were obtained with the two HARPS spectrometers, ESO/HARPS[35] at the 3.6-m ESO telescope at La Silla Observatory and HARPS-N[36] at the 3.5-m Telescopio Nazionale Galileo in La Palma. These are high-resolution echelle spectrometers optimized for precision radial velocities covering a wavelength range 3800–6800 Å. High stability is achieved by keeping the instrument thermally and mechanically isolated from the environment. All observations were wavelength-calibrated with emission lines of a hollow-cathode lamp and reduced with the Data Reduction Software (DRS). For the ESO/HARPS instrument, two distinct datasets are considered (HARPSpre, HARPSpost) corresponding to data acquired before and after a fibre upgrade that took place in June 2015. Radial velocities were obtained using the TERRA[37] software, which builds a high signal-to-noise template by co-adding all the existing observations and then performs a maximum likelihood fit of each observed spectrum against the template yielding a measure of the Doppler shift and its uncertainty. The analysis of the initial HARPSpre dataset, which spans about 6 years, revealed a very prominent signal at a period compatible with 1 year. Thorough investigation led to the conclusion that this is a spurious periodicity caused by the displacement of the stellar spectrum on the detector over the year and the existence of physical discontinuities in the detector structure[38]. We calculated new velocities by removing an interval of ±45 km $s^{-1}$ around the detector discontinuities to account for the amplitude of Earth's barycentric motion. After this correction, all search analyses showed the 1-year periodic signal disappearing well below the significance threshold, although some periodicity remains (possibly related to residual systematic effects in all datasets).

We also use radial velocity measurements of Barnard's star obtained with the UVES spectrograph on the 8.2-m VLT UT2 at Paranal Observatory in the context of the M-dwarf programme executed between 2000 and 2008 (ref. 4). New radial velocity measurements were obtained by reprocessing the iodine-based observations as in ref. 10 using up-to-date reduction codes as those used in the HIRES, PFS, and APF spectrometers.

Barnard's star was observed almost daily in the context of the *CARMENES survey of rocky planets around red dwarfs*[39], which employs the CARMENES instrument, a stabilized visible and NIR spectrometer on the 3.5-m telescope of Calar Alto Observatory. The data were pipeline-processed and radial velocities and their uncertainties were measured with the SERVAL algorithm[40], which is based on a template-matching scheme. For this study we employed visual-channel radial velocities, which correspond to a wavelength interval 5200–9600 Å. Because of instrument effects, data are further corrected by calculating a night-to-night offset (generally below 3 m $s^{-1}$) and a nightly slope (less than 3 m $s^{-1}$ peak to peak) from a large sample of observed stars. Barnard's star was excluded from the calibration to avoid biasing the results. The origin of the offsets is still unclear but they are probably related to systematics in the wavelength solution, light scrambling, and a slow drift in the calibration source during the night. After the corrections, CARMENES data have similar precision and accuracy to those from ESO/HARPS[41].

## Barycentric correction, secular acceleration and other geometric effects

Although stellar motions on the celestial sphere are generally small, the measurement of precision radial velocities must carefully account for some perspective effects, including both the motion of the target star and the observer. This includes, in particular, secular acceleration[4]. A thorough description of a complete barycentric correction scheme down to a precision of <1 cm $s^{-1}$ is given in ref. 42. We ensured that the barycentric corrections employed in all our datasets agree with the code written by ref. 42. Given its proximity to the Sun and high proper motion, Barnard's star is particularly susceptible to errors due to unaccounted terms in its motion. We systematically revised the apparent Doppler shifts accounting for the small but significant changes in the apparent position over time.

Uncertainties in the astrometry (parallax, radial velocity, and proper motion) could propagate into small residual signals in the barycentric correction. We performed numerical experiments to assess the impact of such uncertainties. ED Figure 3 shows the spurious one-year signal expected by introducing a shift of 150 mas (10 times larger than the uncertainties in the Hipparcos catalogue) in both right ascension (R.A.) and declination (Dec.) over a time-interval between years 2000 and 2018. The peak-to-peak amplitudes for such errors are of the order of 4 cm $s^{-1}$. The next larger terms are those that couple the proper motion with the tangential velocity of the star and the tangential velocity of the observer. For this experiment we introduced errors of 15 mas $yr^{-1}$ in both proper motions in the direction of increasing R.A. and Dec., and 15 mas in the parallax (10 times larger than the uncertainties in the Hipparcos catalogue). The spurious signals caused by proper motion contain a trend (change in secular acceleration) and signal with a period of 1 yr growing in amplitude with time. The 1-yr periodicities are rather small and not significant, but the secular trend can produce detectable effects mostly due to the error in the parallax. The effect of errors at 1, 3 and 10σ level of HIPPARCOS uncertainties are shown in the bottom panel of ED Figure 3. Crucially, this signal consists of a trend which is easily included in the model without any major impact on the significance of the planet candidate signal.

# Models and statistical tools

**Doppler model.** The Doppler measurements are modeled using the following equations:

$$v(t_i) = \gamma_{INS} + S \cdot (t_i - t_0) + \sum_{p=1}^{n} f_p(t_i)$$

$$f_p(t_i) = K_p \cos\left[\nu_p\left(t_i;\ P_p, M_{0,p}, e_p\right) + \varpi_p\right] + e_p \cos \varpi_p,$$

where $\gamma_{INS}$ (constant offset of each instrument) and $S$ (linear trend) are free parameters. All signals are included in the Keplerian $f_p$, and for each planet is the Doppler semi-amplitude, $P_p$ is the orbital period, $M_{0,p}$ is the mean anomaly at $t_0$, $e_p$ is the orbital eccentricity and $\varpi_p$ is the argument of periastron of the orbit. Precise definitions of the parameters and the calculation of the true anomaly $\nu_p$ can be found in, e.g., ref. 48. In some cases, the orbits are assumed to be circular and the *Keplerian* term simplifies to:

$$f_{p,circ}(t_i) = K_p \cos\left[\frac{2\pi}{P_p}(t_i - t_0) + M_{0,p}\right]$$

which only has three free parameters ($K_p$, $P_p$, and $M_{0,p}$). This model is used in initial exploratory searches or when analyzing time-series that do not necessarily contain Keplerian signals (e.g., activity proxies).

**Statistical figure-of-merit.** The fit to the data is obtained by finding the set of parameters that maximize the *Likelihood function, L*, which is the probability distribution of the data fitting the model. $L$ can take slightly different forms depending on the noise model adopted. For measurements with normally distributed noise it can be written as

$$L = \frac{1}{(2\pi)^{\frac{N_{obs}}{2}}} |C|^{-1/2}\, exp\left[-\frac{1}{2}\sum_{i=1}^{N_{obs}}\sum_{j=1}^{N_{obs}} r_i r_j C_{ij}^{-1}\right]$$

$$r_i = v_{i,obs} - v(t_i)$$

Where $r_i$ is the residual of each observation $i$, $C_{ij}$ are the components of the covariance matrix between measurements, and $|C|$ is its determinant. Starting from this definition, there are three types of models that we consider.

**White noise model (W)**. If all observations are statistically independent from each other, all variability is included in $v(t_i)$ and the covariance matrix is diagonal. In this case, the logarithm of $L$ simplifies to:

$$\ln L_W = -\frac{N_{obs}}{2}\ln 2\pi - \frac{1}{2}\sum_{i=1}^{N_{obs}} \ln(\epsilon_i^2 + s_{INS}^2) - \frac{1}{2}\sum_{i=1}^{N_{obs}} \frac{r_i^2}{\epsilon_i^2 + s_{INS}^2} ,$$

where $\epsilon_i$ is the nominal uncertainty of each measurement, and $s_{INS}$ is an excess noise component (often called jitter parameter) for each instrument. We call this model, the white noise model (W) as it implicitly assumes that the noise has a uniform power distribution in frequency space.

**Moving average (MA).** Auto-Regressive Moving Average models can also be used (ARMA, ref. 49) when measurements depend on the previous ones in a way that is difficult to parameterize with deterministic functions (e.g., quasi-periodic variability, Brownian motion, impulsive events, etc.). In our case, we use an ARMA model only containing one Moving Average (MA) term assuming that each measurement is related to the previous residual as

$$r_{i,MA} = v_{i,obs} - \left(v(t_i) + r_{i-1,MA}\alpha_{INS}e^{-(t_i-t_{i-1})/\tau_{INS}}\right) .$$

This model contains two additional parameters for each instrument: the coefficient $\alpha_{INS}$ and the time-scale $\tau_{INS}$, representing the strength and time-coherence of the correlated noise, respectively[50].

**Gaussian Process (GP)**. Finally, the most general model, often called Gaussian Processes (or GP), consist of parameterizing the covariance matrix[51], and can be generally written as:

$$C_{ij}^2 = s_{INS}^2\delta_{ij} + \kappa(\tau_{ij})$$

where $\kappa$ is the so-called kernel function, and it is a function of the time difference between observations $\tau_{ij} \equiv |t_i - t_j|$ and some other free parameters. Many kernel functions exist with different properties. Here we consider the stochastically-driven damped Simple Harmonic Oscillator[52] (SHO), which has the form:

$$\kappa(\tau) = C_0 e^{-\tau/P_{\rm life}} \begin{cases} \cosh\left(\eta\frac{2\pi\tau}{P_{\rm rot}}\right) + \frac{P_{\rm rot}}{2\pi\eta P_{\rm life}}\sinh\left(\eta\frac{2\pi\tau}{P_{\rm rot}}\right), P_{\rm rot} > 2\pi P_{\rm life} \\ 2\left(1+\frac{2\pi\tau}{P_{\rm rot}}\right), P_{\rm rot} = 2\pi P_{\rm life} \\ \cos\left(\eta\frac{2\pi\tau}{P_{\rm rot}}\right) + \frac{P_{\rm rot}}{2\pi\eta P_{\rm life}}\sin\left(\eta\frac{2\pi\tau}{P_{\rm rot}}\right), P_{\rm rot} < 2\pi P_{\rm life} \end{cases}$$

Where $P_{\rm rot}$ is the stellar rotation period, $P_{\rm life}$ is the lifetime of active regions, $C_0$ is a scaling factor proportional to the fraction of stellar surface covered by active regions, and $\eta = |1 - (2\pi P_{\rm life}/P_{\rm rot})^{-2}|^{1/2}$. This model is popular in astrophysical applications because its three parameters can be associated to physical properties.

**False Alarm Probability (or p-value)**. We use the frequentist concept of False Alarm Probability of detection (FAP hereafter) to assess statistical significance. FAP is formally equivalent to the so-called p-value used in other applications. The statistical significance of the detection of a planet is a problem of null hypothesis significance test, where the null hypothesis is a model with *n* signals (null model), and the model to be benchmarked contains $n$+1 signals with a correspondingly larger number of parameters. The procedure is as follows:

i. We start computing ln $L$ of the null model, containing all *n* detected signals and nuisance parameters (jitters, trend, etc.)
ii. Next, ln $L$ is maximized by adjusting all the model parameters together with the parameters of a sinusoid for a list of test periods for signal $n$+1. Then, the logarithm of the improvement of the likelihood function with respect to the null model is computed ($\Delta$ ln $L_{P,n+1}$) at each test period $P$ and plotted against each other generating a so-called log-likelihood periodogram[53].
iii. The largest $\Delta$ ln $L_{P,n+1}$ (peak in periodogram) indicates the most favoured period for the new signal. This value is then compared with the probability of randomly finding such an improvement when the null hypothesis is true, which is the desired FAP[54]. A FAP around 1% would be considered tentative evidence, and below $10^{-3}$ (or 0.1%) is considered statistically significant.

All FAP assessments and significances presented in this work, including Doppler data and activity indicators, are computed using this procedure. We note that FAPs will depend on the adopted model (W, MA or GP).

**Bayesian tools and analyses**. We also applied Bayesian criteria to the detection of signals (Bayesian factors as in ref. 14), but these lead to conclusions and discussions qualitatively similar to those presented, so they are omitted for brevity.

Concerning median values and credibility intervals presented in tables, a standard custom-made code implementing a Markov Chain Monte Carlo (MCMC) algorithm as presented in ref. 55. In all the cases, uniform priors in all the parameters were assumed, with the exception of the periods. In that case, the prior was chosen to be uniform in frequency and an upper limit to the period was set to twice the timespan of the longest dataset (~12 000 days).

## Noise models and experiments applied to our datasets

If the presence of spurious Doppler variability caused by stellar activity is suspected, checking the significance of the detections under different assumptions about the noise is advisable[56]. The significance assessments in the main manuscript are given assuming an MA model for the radial velocity analyses, and W models for all other sets (photometry, activity indices). This section provides the justification for such assumption. White noise models are good for preliminary assessments but they are prone to false positives[14]. On the other hand, GPs tend to produce overly conservative significance assessments leading to false negatives.

We investigated the performances of the different noise models (W, MA, and GP) by analysing the combination of three datasets in more detail: HIRES, HARPSpre and CARMENES. These are the relevant ones because they contribute most decisively to the improvement of the likelihood statistic (largest number of points, widest timespan, and higher precision). The W model found the signal at $P$ = 233 days with $\Delta \ln L = 42$ (FAP~$3.3 \cdot 10^{-14}$), and the MA model yielded a detection with $\Delta \ln L = 22.3$ (FAP~ $8.6 \cdot 10^{-6}$). On the other hand, a GP using the SHO kernel, yielded a detection with only $\Delta \ln L = 11.6$ (FAP~27%). Despite this rather poor significance, GPs account for *all* covariances including those produced from real signals, which prompted us to carry out a deeper assessment.

We performed simulations by injecting a signal at 233 days (1.2 m $s^{-1}$) and attempted the detection using W, MA and GP models. We first generated a synthetic sinusoidal signal (no eccentricity) and sampled it at the observing dates of the three sets. Random white noise errors were then associated to each measurement in accordance to their formal uncertainties and jitter estimates of each set. When using W and MA models, a one-planet search trivially detected the signal at 233 days yielding $\Delta \ln L = 43$ (FAP ~$1.22 \cdot 10^{-14}$) and $\Delta \ln L = 32$ (FAP~$6.3 \cdot 10^{-10}$) respectively, indicating high statistical significance. On the other hand, adding one planet when using GPs led to a $\Delta \ln L =$ 14 (FAP ~2.7%), indicating that an unconstrained GP (all parameters free) absorbed $\Delta \ln L \sim 29$, even in the absence of any *true* correlated noise. This reduction is comparable to that observed in the real dataset (from $\Delta \ln L = 42$ for the W model, to $\Delta \ln L = 11.6$ when employing a GP model as discussed earlier), supporting the hypothesis that the GP is substantially absorbing the real signal, even if its parameters are set to match the rotation period of the star derived from spectroscopic indices and photometry (see ED Figure 4, for a visual representation of the effect).

The filtering properties of GPs can be better understood in Fourier space (frequency domain). As discussed in ref. 52, GPs fit for covariances within a range of frequencies filtered by the power spectral distribution (PSD) of the kernel function used. In particular, for an SHO kernel, the PSD is centred at the frequency of the oscillator, $\nu = 2\pi/P_{\rm rot}$, and has full-width-at-half-maximum $2/P_{\rm life}$. The activity indices of Barnard's star imply that $\nu$ and $2/P_{\rm life}$ are comparable and of the order of $10^{-2}$ day$^{-1}$. Consequently, the GP strongly absorbs power (i.e., $\Delta \ln L$) from signals in a frequency range $10^{-2} \pm 10^{-2}$ day$^{-1}$, which spans periods from 50 days to infinity as illustrated by the black line in ED Figure 4. Most of the proposed kernels in the literature are very similar to the SHO kernel, so similar filtering properties are to be expected.

In a separate set of simulations, we checked the sensitivity of W, MA and GP models to false positives by creating synthetic data generated from covariances. The results were in general agreement with ref. 14 in the sense that the MA models have best statistical power. Furthermore, 300 000 data sets were generated using the MCMC sampling of the SHO parameters. $P_{\rm rot}$ and $P_{\rm life}$ pairs were derived from MCMC fits to the H$\alpha$ time series and the corresponding $C_0$ parameters were obtained from an empirical relationship obtained from fitting GP kernels with fixed $P_{\rm rot}$ and $P_{\rm life}$ to our real RV datasets. Next, synthetic observations were obtained using a multivariate random number generator from the covariance matrix for all the epochs. Reported uncertainties and jitter estimates for each observational dataset were added in quadrature and consistent white noise was also injected. Finally, a synthetic set was only accepted if having a root mean square within 0.1 m s$^{-1}$ of the real value. We then performed a maximum likelihood search using the MA model, and the solution with maximum likelihood was recorded in each case. This process produced a distribution of false alarms as a function of $\Delta \ln L$ and $P_{\rm rot}$ as illustrated in ED Figure 5. This leads to a FAP ~ 0.8% for our candidate signal. Although this is not an extremely low value, we consider it sufficiently small to claim a detection given that we followed a rather conservative procedure, and given the existing degeneracies between signals and correlated noise models. Crucially, if we had carried out a deep scrutiny of each of the false alarms as we did with our real dataset we would have discarded the fraction failing our sanity checks (steady growth in signal strength, existence of a significant signal in populated dataset pairs, consistent offsets in overlapping regions, etc). This would reduce the numerical value of the estimated FAP using this procedure.

In summary, we find that the most adequate models to account for the noise and maximize the detection efficiency in this period domain are those using MA terms, and that the 233-day signal is statistically significant under these models.

## Zero-points between datasets

Calculation of the zero-points between the different datasets is a key element to ensure unbiased results, detection of genuine signals and to avoid introducing spurious effects. The best-fitting model is a self-consistent fit of the datasets allowing for a variable zero-point offset that is optimized via maximum likelihood together with the search for periodicities. To validate these results, we used a complementary approach based on searching for overlapping coverage (typically a few nights) between different datasets to calculate average differences and thus directly measure zero-point offsets. We worked recursively, piecing datasets together one by one depending on the existence and size of overlap regions. We optimised the averaging window and selected that providing the best agreement in a 3-way comparison. This is a trade-off between window size, number of points, and measurement error. Periods below the window duration are affected by this process but our focus lies in a period of 233 days. Any window size smaller than a few tens of days does not impact the results.

The window parameters and the differences between the manually-computed zero-point offsets and the values resulting from the optimization routine (considering a long-period signal) are given in ED Table 3. The compatibility of the zero-points calculated using two completely independent methods is very good. Only for UVES does a difference significantly larger than 1 sigma appear. This can be attributed to the sparse sampling of the observations leading to small overlap between the datasets. Also, the zero-point is based on a few measurements from HIRES that appear to deviate systematically from the average. Because of the reduced overlap, the resulting zero-point is critically dependent on the window size and thus is quite unreliable. The most populated datasets (HIRES, HARPSpre and CARMENES) have excellent zero-point consistency. Additionally, the agreement of the general offsets of the combined Set1 (HIRES, UVES, HARPSpre, APF, PFS) and Set 2 (CARMENES, HARPS-N, HARPSpost) is remarkable (ED Table 3). This is related to the presence of the long-term signal, which is found naturally when calculating manual offsets and confirmed from the global optimization including a long-period prior.

# Stellar activity analysis

Barnard's star is considered to be an aged, inactive star, but it appears to have small changing spots that make its rotation period tricky to ascertain. Spectroscopic indices (Hα and Ca II H&K) and photometric measurements were used to estimate the period range in which signals from stellar activity are present. In all cases, the modelling of the data was performed using the same methodology as for the radial velocities, including the optimization of zero-point offsets and jitter terms for the different instruments, but assuming sinusoidal signals (zero eccentricity). As a result of the analysis, the stellar rotation period can be constrained to be in the range 130–150 d from all indicators, and there is also evidence for long-period modulation, which could be related to an activity cycle. No significant variability related to magnetic activity is present around 233 days, where the main radial velocity periodic signal is found. A thorough review and analysis of all data on activity indicators for Barnard's star will be presented in a separate publication.

**Spectroscopy - Hα index.** Stellar activity was studied using the available spectroscopic data on Barnard's star. The Hα index was calculated using three narrow spectral ranges covering the full Hα line profile and two regions on the pseudo-continuum at both sides of the line, after normalizing the spectral order with a linear fit[3]. The error bars were estimated by adopting the standard deviation of the fluxes in a small local continuum region close the core of the lines as the uncertainty of the individual fluxes. The Hα index was measured in 618 night-averaged spectra acquired with seven different instruments covering a timespan of 14.5 years. The analysis of the resulting time series (Figure 2) yields a high-significance (FAP<<0.1%) periodic signal at 133 days, and a second also highly-significant signal at 191 days. We interpret the 133-day periodicity as tracing the stellar rotation period. This value is in relatively good agreement with a previous determination of 148 d (ref. 3). The longer period signal could be a consequence of the non-sinusoidal nature of the variability, the finite lifetime of active regions or the presence of differential rotation. The analysis of the $H\alpha$ index does not reveal any significant long-term (P > 1000 d) modulation.

**Spectroscopy – S-index.** The S-index[43] derived from the Ca II H&K lines was only available for five instruments (APF, HARPS-N, HARPSpost, HARPSpre, and HIRES). The S-index was estimated from 384 night-averaged spectra covering a similar time span as $H\alpha$. Two long-period signals were extracted from the analysis of the time series (Figure 2) at periods of 4300 days and 560 days. The next strongest significant signal, with FAP~$10^{-4}$, has a period of 143 days, and it is probably associated with stellar rotation. Using an empirical relationship[44], the activity-induced RV signal corresponding to this rotation period is predicted to be ~0.6 m $s^{-1}$. The long-term signal found from the S-index is consistent with estimates of activity cycles from photometric time series in other M stars of similar activity levels[45].

**Photometry.** Photometry from the literature includes data from the All Sky Automated Survey (ASAS, ref. 46) and the MEarth Project[47] database. We also used unpublished photometry from the 0.8-m Four College Automated Photoelectric Telescope (FCAPT, Fairborn Observatory, Arizona, USA) and the 1.3-m Robotically-Controlled Telescope (RCT, Kitt Peak National Observatory, Arizona, USA). In addition, new observations were acquired within the RedDots2017 campaign (https://reddots.space/) from the following facilities: the 0.90-m telescope at Sierra Nevada Observatory (Granada, Spain), the robotic 0.8-m Joan Oró telescope (TJO, Montsec Astronomical Observatory, Lleida, Spain), Las Cumbres Observatory network with the 0.4-m telescopes located in Siding Spring Observatory, Teide Observatory and Haleakala Observatory, the ASH2 0.40-m robotic telescope at San Pedro de Atacama (Celestial Explorations Observatory, SPACEOBS, Chile), and from 14 observers of the American Association of Variable Stars Observers (AAVSO). A comprehensive summary of these measurements and contributors will be given in a paper in preparation. The data cover about 15.1 years of observations with 1634 epochs, an rms of 13.6 mmag and a mean error of 9.8 mmag. The analysis of the combined datasets (Figure 2) indicates long-term modulations of 4500 days and 1300 days (semi-amplitudes of 10 and 5 mmag, respectively) and two significant periods at 144 days and 201 days (semi amplitudes of ~3 mmag). The interpretation is that the long-term modulation may be caused by an activity cycle while the signals at 144 days and 201 days are likely related to the base stellar rotation period and to the effects of the finite lifetime of active regions and differential rotation at different latitudes. The resulting periods are consistent with the results from the spectroscopic indices. A rotation period of 130.4 days and ~5 mmag semi-amplitude had been previously reported[13] from photometric observations albeit with low significance (FAP~10%).

**Code availability.** The SERVAL template-matching radial velocity measurement tool used for CARMENES data can be found at https://github.com/mzechmeister/serval. The TERRA template-matching radial velocity measurement tool and various custom periodogram analysis and MCMC tools are codes written in Java by G. Anglada-Escudé and are available upon request. Other public codes and facilities used to model the data include GLS (http://www.astro.physik.uni-goettingen.de/~zechmeister/gls.php), Systemic Console (https://github.com/stefano-

meschiari/Systemic-Live), Agatha (https://github.com/phillippro/agatha), Celerite (https://github.com/dfm/celerite.git) and EMCEE (https://github.com/dfm/emcee).

**Data availability.** The public high-resolution spectroscopic raw data used in the study can be freely downloaded from the corresponding facility archives (http://koa.ipac.caltech.edu for HIRES; http://archive.eso.org for UVES, HARPSpre and HARPSpost; http://archives.ia2.inaf.it/tng for HARPS-N, and https://mthamilton.ucolick.org/data for APF). Proprietary raw data are available from I.R. on reasonable request. The nightly-averaged, fully-calibrated radial velocities, spectroscopic indices and photometric measurements are available online.

| Instrument | Calib. method | Time | Epochs | Program ID | PI/Group |
|---|---|---|---|---|---|
| Keck/HIRES | Iodine | 06/1997–08/2013 | 186 | † | Vogt, Butler, Marcy, Fischer, Borucki, Lissauer, Johnson (and several more with <10 obs) |
| VLT/UVES | Iodine | 04/2000–10/2006 | 75 | 65.L-0428<br>66.C-0446<br>267.C-5700<br>68.C-0415<br>69.C-0722<br>70.C-0044<br>71.C-0498<br>072.C0495<br>173.C-0606<br>078.C-0829 | UVES survey; Kürster |
| ESO/HARPSpre | Hollow-cathode lamp | 04/2007–05/2013 | 118 | 072.C-0488<br>183.C-0437<br>191.C-0505 | Mayor, Bonfils, Anglada-Escudé |
| Magellan/PFS | Iodine | 08/2011–08/2016 | 39 | Carnegie-California survey | Crane, Butler, Shectman, Thompson |
| APF | Iodine | 07/2013–07/2016 | 43 | LCES/APF planet survey | Vogt, Butler (and several programmes) |
| HARPS-N | Hollow-cathode lamp | 07/2014–10/2017 | 40 | CAT14A_43<br>A27CAT_83<br>CAT13B_136<br>CAT16A_99<br>CAT16A_109<br>CAT17A_38<br>CAT17A_58<br>CAT17B_140 | Amado, Rebolo, González Hernández, Berdiñas |
| CARMENES | Hollow-cathode lamp | 02/2016–11/2017 | 201 | CARMENES GTO survey | CARMENES consortium |
| ESO/HARPSpost | Hollow-cathode lamp | 07/2017–09/2017 | 69 | 099.C-0880 | Anglada-Escudé/RedDots |

**Extended Data Table 1: Log of observations of Barnard's star with seven different facilities.** In the case of ESO/HARPS, the "pre" and "post" tags indicate data obtained before and after a hardware upgrade in June 2015. A secular acceleration term of 4.497 m $s^{-1}$ $yr^{-1}$ due to change in perspective over time (ref. 4) was removed from all the sets when applying the barycentric correction to the raw Doppler measurements. †: H7aH, K01H, N02H, N03H, N05H, N06H, N10H, N12H, N14H, N15H, N19H, N20H, N22H, N24H, N28H, N31H, N50H, N59H, U01H, U05H, U07H, U08H, U10H, U11H, U12H, U66H, H38bH, A264Hr, A285Hr, A288Hr, C110Hr, C168Hr, C169Hr, C199Hr, C202Hr, C205Hr, C232Hr, C240Hr, C275Hr, C332Hr, H174Hr, H218Hr, H238Hr, H244Hr, H257Hr, H305Hr, N007Hr, N014Hr, N023Hr, N024Hr, N054Hr, N085Hr, N086Hr, N095Hr, N108Hr, N118Hr, N125Hr, N129Hr, N131Hr, N134Hr, N136Hr, N141Hr, N145Hr, N148Hr, N157Hr, N168Hr, U009Hr, U014Hr, U023Hr, U026Hr, U027Hr, U030Hr, U052Hr, U058Hr, U064Hr, U077Hr, U078Hr, U082Hr, U084Hr, U115Hr, U131Hr, U142Hr, Y013Hr, Y065Hr, Y292Hr

| Dataset | Jitter (m s$^{-1}$) | γ (m s$^{-1}$) |
|---|---|---|
| Keck/HIRES | $2.28^{+0.19}_{-0.18}$ | $1.26^{+0.38}_{-0.32}$ |
| VLT/UVES | $2.42^{+0.25}_{-0.22}$ | $3.83^{+0.58}_{-0.57}$ |
| ESO/HARPSpre | 0.92±0.14 | $0.97^{+0.36}_{-0.27}$ |
| Magellan/PFS | $0.96^{+0.37}_{-0.41}$ | $1.76^{+0.41}_{-0.37}$ |
| APF | $2.78^{+0.51}_{-0.44}$ | $2.16^{+0.65}_{-0.63}$ |
| HARPS-N | $1.45^{+0.27}_{-0.23}$ | $1.37^{+0.65}_{-0.63}$ |
| CARMENES | $1.76^{+0.15}_{-0.14}$ | 1.55±0.65 |
| ESO/HARPSpost | $1.16^{+0.19}_{-0.18}$ | 1.46±0.69 |

**Extended Data Table 2: Additional fit parameters and fit results.** The individual zero-points and jitter terms are optimized for each dataset by maximizing the likelihood function. The model included a global time trend that results in a best-fitting value of +0.33±0.07 m s$^{-1}$ yr$^{-1}$. It should be noted that the original individual datasets were previously shifted to have null relative offsets in the overlapping regions (see ED Table 3) and referred to the zero-point level of the Keck/HIRES dataset. This implies that the optimized γ parameters in the table are not totally arbitrary but expected to be relatively similar. The parameters and their uncertainties are determined by calculating the median values and 68% credibility intervals of the distribution resulting from the MCMC run.

| Datasets | Window size (±days) | Measur. used | Diff. (manual-optimized) (m s$^{-1}$) |
|---|---|---|---|
| UVES–HIRES | 10 | 28 | 2.53±0.65 |
| HARPSpre–HIRES | 10 | 291 | −0.29±0.31 |
| PFS–HIRES | 10 | 130 | 0.49±0.49 |
| APF–PFS | 10 | 17 | 0.37±0.85 |
| HARPSpost–CARMENES | 2 | 161 | −0.09±0.33 |
| HARPS-N–CARMENES | 2 | 75 | −0.18±0.39 |
| Set1–Set2 | 8 | 14 | −0.24±0.52 |

**Extended Data Table 3: Zero-point offsets between overlapping radial velocity datasets from different instruments.** Manual offsets are calculated from common regions of pairs of datasets for window sizes selected to ensure sufficient statistics and consistency in the cases of 3-way overlap. The last column lists the difference between the zero-points calculated manually and those resulting from the global optimization, showing general good agreement (values compatible with zero), except for the UVES dataset. Also, two distinct time regions are identified in the data and can be compared. Set 1 includes data from HIRES, UVES, HARPSpre, APF and PFS. Set 2 contains data from CARMENES, HARPS-N and HARPSpost. The relative zero-point between these two sets is poorly defined because of very limited overlap but the consistency between the manual and optimized values is found to be very good. All error bars correspond to 1σ values.

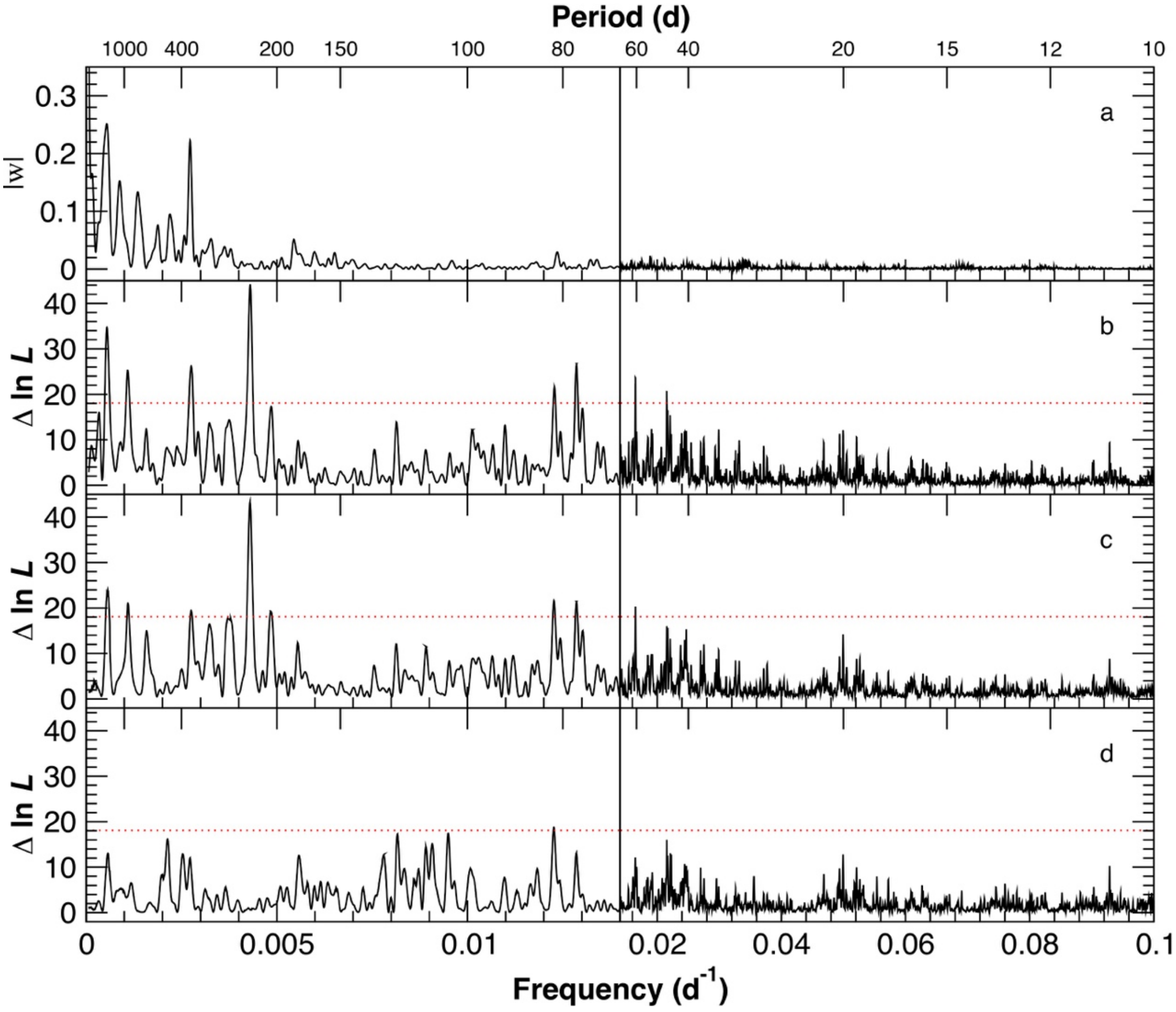

**Extended Data Figure 1: Hierarchical periodogram analysis.** Panel a shows the magnitude of the window function of the combined datasets. The rest of the panels show the likelihood periodogram of the radial velocity measurements considering, subsequently, first signal search (panel b), the residuals after modelling a long-period (6600 days) signal as explained in the text (panel c) and the residuals after modelling long-period and 233-day periodicities (panel d). No high-significance signals are left, in particular in the 10–40-day region, corresponding to the conservative habitable zone. The region below 10 days is not shown for clarity, but it is also devoid of significant periodic signals down to the Nyquist frequency of the dataset (2 days). Two different scales for the horizontal axis are used to improve visibility of the low frequency range.

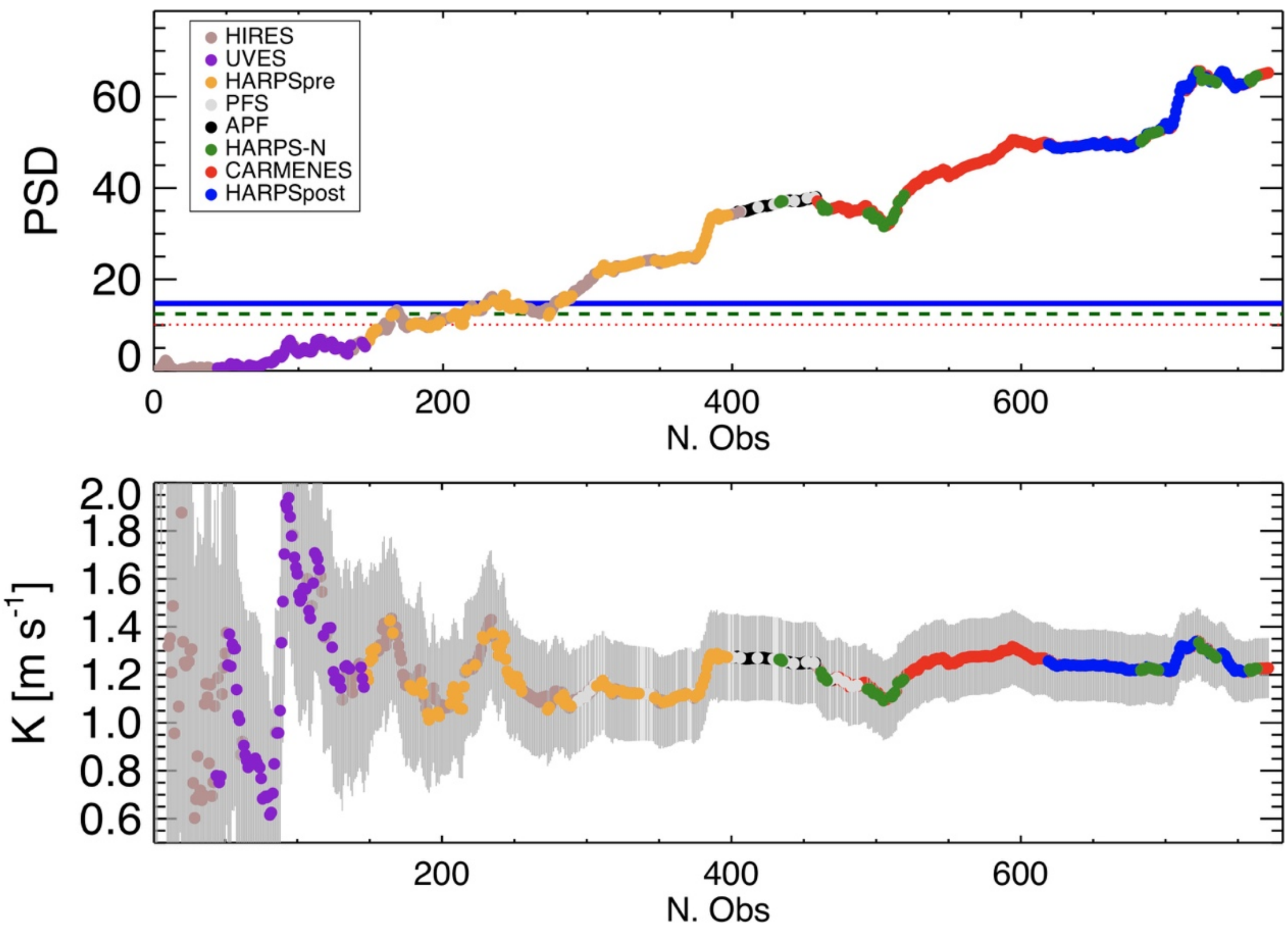

**Extended Data Figure 2: Evolution of the significance of the 233-day signal.** The top panel shows the power spectral density[57] (PSD) of a stacked periodogram[58,59] while the bottom panel depicts a cumulative measurement of the semi-amplitude of the signal. The horizontal red dotted line, green dashed line, and blue solid line show the 10%, 1% and 0.1% FAP thresholds. The evolution of the significance is stable with time and the variations in the amplitude over the last 9 years of observations are smaller than 5% of the measured amplitude. Both the steady increase in signal significance and the stable amplitude are consistent with the expected evolution of the evidence for a signal of Keplerian origin.

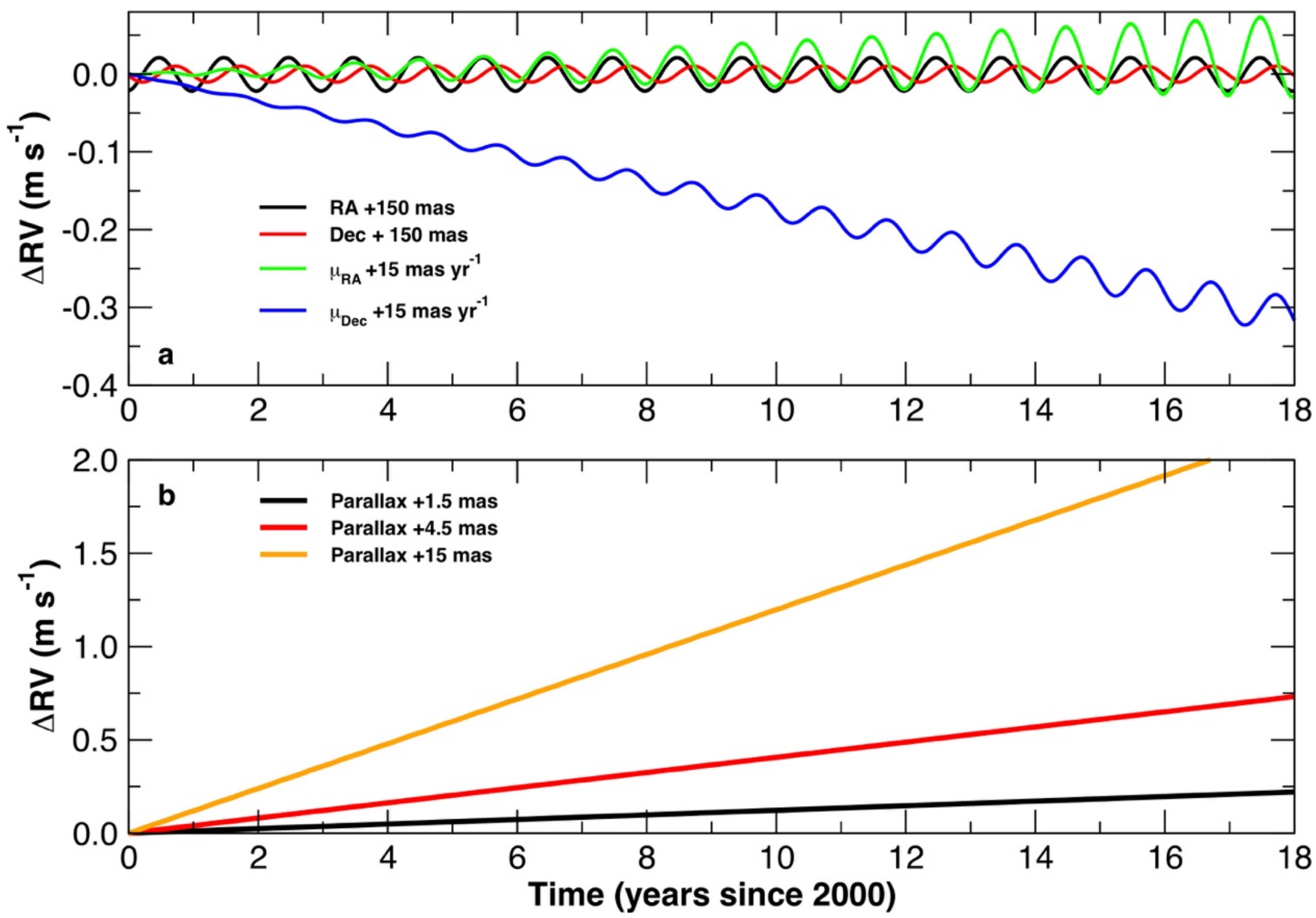

**Extended Data Figure 3: Propagation of astrometric errors to radial velocity systematics.** Panel a shows the spurious radial velocity effect that would be caused by offsets with respect to the catalogue coordinates (black and red) and proper motions (green and blue). Panel b illustrates the radial velocity effect caused by an offset in the parallax with respect to the catalogue value. The uncertainties of the astrometric parameters for Barnard's star from the Hipparcos catalogue were used in the barycentric corrections, and they are approximately 10 times smaller than the values used in this plot (15 mas in position, 1.5 mas $yr^{-1}$ in proper motion, and 1.5 mas in parallax), implying that catalogue errors introduce undetectable signals.

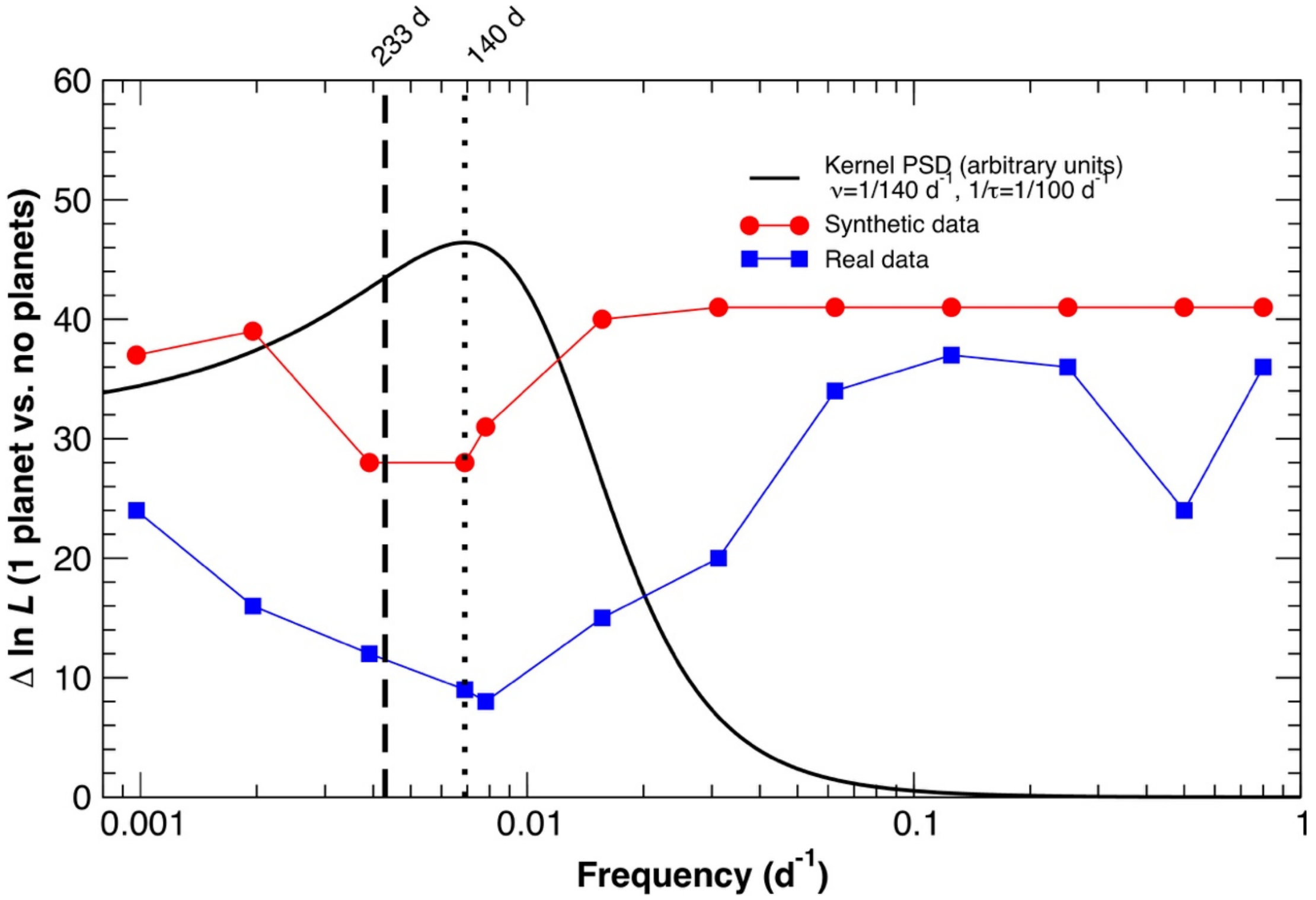

**Extended Data Figure 4: Effect of Gaussian Processes (GP) modelling when applied to synthetic and real data.** Blue squares represent the improvement of the ln-likelihood using a GP to model the correlated noise when trying to detect a first signal. The same procedure is applied to simulated observations generated with white noise and a sinusoidal signal consistent with the planet candidate parameters (red circles). Even in absence of true correlated noise, the GP absorbs a substantial amount of significance ($\Delta \ln L \sim 30$ for this selection of kernel parameters). The adopted kernel is a damped stochastic harmonic oscillator (SHO), with a damping timescale of $\tau = P_{\rm life} = 100$ days and each point corresponds to different values for the oscillator frequency ν (x-axis). The power spectral distribution (PSD) of an SHO kernel with $\nu = 140^{-1}$ day$^{-1}$ and $\tau = 100$ days is depicted as a black line. The greater reduction in significance occurs when the trial frequency approaches that of the oscillator, but this reduction in significance extends out to a broad range of frequencies therefore acting as a filter. The planet candidate period is marked with a vertical dashed line, and the likely rotation period derived from stellar activity is marked with a vertical dotted line.

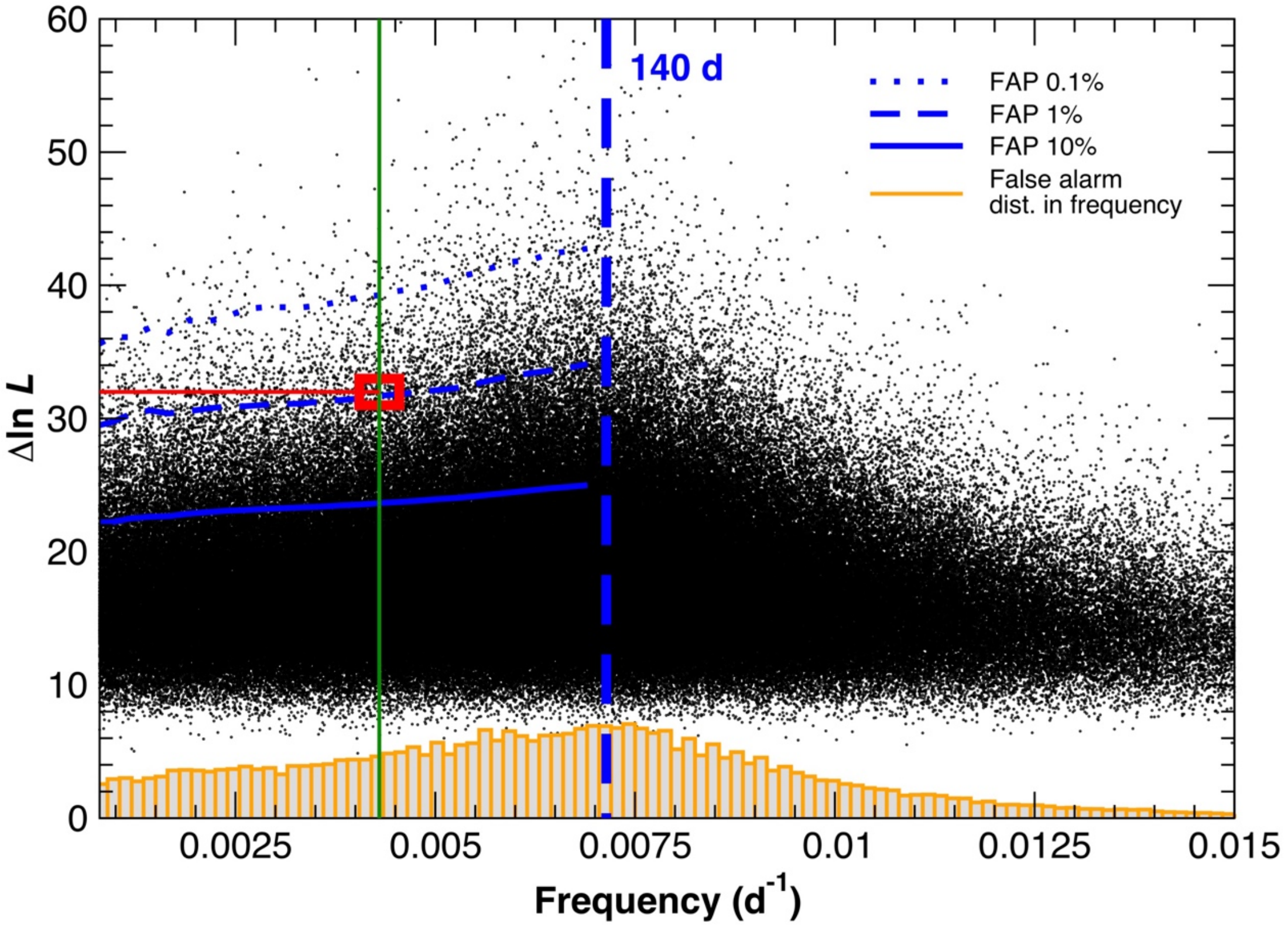

**Extended Data Figure 5: Distribution of empirical false alarms from synthetic observations with correlated noise.** These simulations were obtained by generating synthetic observations following kernels derived from the observations, and then fitted to MA models. The resulting distribution of false alarms shows a clear excess around the measured rotation period of the star (vertical dashed blue line), and at low frequencies (long periods) due to the use of the free offsets in the model (left of the rotation period). The empirical FAP was computed by counting the number of false alarms in the interval $\Delta \ln L \in [32,\infty]$ and frequency $\in [0,1/230]$ (left of the green line and above the red line) and dividing it by total number of false alarms in the same frequency interval (left of the green line). Empirical FAP threshold lines of 10%, 1% and 0.1% are shown for reference. The candidate signal under discussion is shown as a red square and has an empirical FAP of ~0.8%. The orange histogram at the bottom shows the distribution of false alarms in frequency (arbitrary normalization).